\documentclass[12pt]{iopart}

\usepackage[backend=biber,sorting=none,style=numeric-comp]{biblatex}
\usepackage{mwe}
\usepackage{subcaption}

\usepackage[font=footnotesize
,labelfont=bf,margin=1cm]{caption}
\usepackage[justification=justified]{caption}

\usepackage{adjustbox}

\usepackage{comment}
\usepackage{siunitx}
\newcolumntype{d}[1]{D{.}{.}{#1}}

\begin{document}

\title[Large-Scale Deep Learning for Multi-Jet Event Classification]{Large-Scale Deep Learning for Multi-Jet Event Classification}
\author{
Jiwoong Kim$^2$,
Dongsung Bae$^3$,
Kihyeon Cho$^1$,
Junghwan Goh$^3$,
Jaegyoon Hahm$^1$,
Taeyoung Hong$^1$,
Soonwook Hwang$^1$,
Minsik Kim$^1$,
Sungwon Kim$^4$,
Tongil Kim$^4$,
Chang-Seong Moon$^2$,
Hunjoo Myung$^1$,
Hokyeong Nam$^2$, 
Changhyun Yoo$^3$,
Hwidong Yoo$^4$,
}

\address{$^1$ Korea Institute of Science and Technology Information, 245, Daehak-ro, Yuseong-gu, Daejeon, Republic of Korea}
\address{$^2$ Department of Physics, Kyungpook National University, 80 Daehakro, Bukgu, Daegu 41566, Republic of Korea}
\address{$^3$ Department of Physics, Kyung Hee University, 26 Kyungheedae-ro, Dongdaemun-gu, Seoul 02447, Republic of Korea}
\address{$^4$ Department of Physics, Yonsei University, 50 Yonsei-ro Seodaemun-gu, Seoul, 03722, Republic of Korea}
\ead{junghwan.goh@cern.ch,csmoon@cern.ch}
\vspace{10pt}
\begin{indented}
\item[]June 2023
\end{indented}

\begin{abstract}
We report the largest scale deep learning with High Performance Computing (HPC) to physics analysis with the CMS simulation data in proton-proton collisions at 13 TeV. We build a Convolutional Neural Network (CNN) model that takes low-level information as images considering the geometry of the CMS detector and use this model to discriminate \textit{R}-parity violating super symmetry (RPV SUSY) events from the background events with inelastic quantum process from the Standard Model (QCD multi-jet). We compare the classification performance of the CNN method with that of the widely used cut-based method. The signal efficiency (and expected significance) of the CNN method is 1.85 (1.2) times higher than that of the cut-based method. To speed-up the training, the model training is conducted using the Nurion HPC system at the Korea Institute of Science and Technology Information, which is equipped with thousands of parallel \texttt{Xeon Phi} CPUs. Notably, our CNN model shows scalability up to 1024 nodes.
\end{abstract}

%
%
%
%
%

\section{Introduction}
The foremost objective of performing various high-energy physics (HEP) experiments is to understand the elementary particles of the universe and the laws that govern their interactions. One of the major interests is to find new phenomena that cannot be explained by the Standard model (SM) \cite{SM}. For conducting HEP experiments, the neoteric instruments -particle accelerators- such as the Large Hadron Collider (LHC) \cite{evans2008lhc} at CERN are primarily employed. At the LHC, proton bunches are accelerated to velocities near the speed of light and collided together with a center-of-mass energy of 13 TeV to explore new physics processes including supersymmetry (SUSY) \cite{freund1988introduction} and dark matter \cite{bertone2005particle} using detectors such as the CMS \cite{adolphi2008cms} and ATLAS \cite{aad2008atlas}. Observing these processes and evaluating their mechanism are essential to obtain a fundamental understanding of the Universe.
  
In this study, we define the signal process as a multiple jets decay process in the \textit{R}-parity violating (RPV) SUSY mode \cite{RPV}. From a theoretical point of view, this model can provide solutions to the unsolved issues in the SM (e.g.\@ dark matter and mass hierarchies). From an experimental point of view, final state of the RPV SUSY process consists of the well-known SM particles.
  
To search for the signal, SM backgrounds that mimic the signal should be well discriminated. However, in a multi-jet process, it is extremely difficult to reduce only the SM background via a simple cut-based method, in which the discriminator is manually investigated. In particular, the SM process containing the multi-jet as the final state (QCD multi-jet) should be considered as a main background because the final states of both RPV SUSY signal and the QCD multi-jet SM background are almost the same with 10 jets. Additionally, the QCD multi-jet process is 300,000 times more likely to be generated in the LHC at the center-of-mass energy of 13 TeV than does the RPV SUSY process.
Finding the signal dominant region from huge number of QCD multi-jet backgrounds using a cut-based method is a challenging task owing to several limitations, primary among them being the low signal efficiency (i.e. the ratio of total data to signal events), which makes it difficult to observe the signal process. Therefore, to improve the signal efficiency, implementing the machine learning (ML) method is essential.
  
The first goal of this study is to apply ML to classify the RPV SUSY signal from the QCD multi-jet background. Previously, the search for RPV-SUSY process using cut-based method was performed by the CMS collaboration \cite{adolphi2008cms}, using 2.7 fb$^{-1}$ of data acquired at a collision energy of 13 TeV. Through this study, gluino masses smaller than 1360 GeV are excluded at a 95\% confidence level \cite{cms2016search}. We use the results of this study as the benchmark and compare the signal efficiency between the ML method and the cut-based method, at the same background rejection point, with the final event selection criteria used in \cite{cms2016search}. We verify that the ML method shows a higher signal efficiency compared to that of the cut-based method reported in \cite{cms2016search}.
  
Reportedly, with the application of ML in HEP, it has been demonstrated that the signal efficiency extracted using the Deep Neural Network (DNN) with low-level features, such as the position of the particle in the detector and its reconstructed momentum, is higher than that obtained using the conventional ML (e.g.\@ boosted decision tree and shallow neural network) with high-level features, which are manually constructed \cite{baldi2014searching, baldi2015enhanced}. According to these studies, the DNN model learns the high-level features directly from the low-level features, thereby showing a higher classification performance than do the conventional method with manually constructed high-level features.

These results reveal that training a more complex DNN model, such as the Convolutional neural network (CNN) \cite{lecun1999object}, with lower-level data like a detector image, instead of the reconstructed physics variables, can open new avenues for high signal efficiency. The CNN is well known to show the highest accuracy in image recognition \cite{Alex}, indicating the possibility of, classifying signal with backgrounds in HEP using a detector images is highly motivated. In particular, the idea can be used in the analysis of physical processes, including a large number of jets in their final state, which are difficult to analyze because of the numerous complex algorithms, such as $\mathrm{k_{t}}$, anti-$\mathrm{k_{t}}$, Cambridge, etc \cite{jetalgoreview}, that are used to define the jet objects. The CNN model directly extracts features of signal and background processes from detector images, thereby avoiding the loss of efficiency due to jet reconstruction.
  
Consequently, we build a CNN model with detector images as the training input data, using an approach similar to that reported in  \cite{de2016jet,metodiev2017classification,bhimji2017deep,andrews2019exploring}. However, we further optimize the data structure and CNN architecture to improve the training performance. First, the transverse momentum weight is applied on the tracker image. Then, the CNN architecture is developed by incorporating the azimuthal symmetry of the image data obtained from the shell of a cylindrical detector. Further, two different pixel-sizes are considered ($64\times64$ and $224\times224$) in the image generation step. Images with a large number of pixel ($224\times224$) is selected because it is expected to have more information than does the $64\times64$ pixel images, according to its higher resolution. These methods increased the training performance, compared to the direct application of the methods used in \cite{de2016jet,metodiev2017classification,bhimji2017deep,andrews2019exploring}.
  
The second goal of this study is to ensure that our CNN architecture runs efficiently in the HPC environment and shows scalability, which is the ability of a system to reduce training time as the number of nodes is increased. With the application of high performance computing (HPC) to DNN, it is possible to minimize the training time effectively through distributed training, which is advantageous for HEP analysis that generally requires extremely long training times. For example, at the LHC, data acquisition, approximately 20 times more than the current integrated data (from 150 to 3000 $\mathrm{fb^{-1}}$), is planned to aid in the discovery of new physical processes~\cite{HLLHC}.
  
The scalable deep learning on HEP analysis was implemented using the ATLAS dataset on the Cori HPC at the National Energy Research Scientific Computing Center, whose major computing resource is predominately composed of Intel Knights Landing (KNL) ``Xeon Phi'' CPUs \cite{kurth2017deep}. We reference this study using CMS simulation data on the Nurion HPC \cite{nurion} system at the Korea Institute of Science and Technology Information (KISTI). Since the Nurion HPC system consists of 8305 \texttt{Xeon Phi} CPUs, we apply distributed training method similar to that used in \cite{kurth2017deep}. In addition, optimization of the training architecture through various methods, such as the application of a new deep learning framework (\textsc{Pytorch}) \cite{paszke2019pytorch} and data loading method are performed to improve the scalability in the Nurion HPC system.
  
This paper is organized as follows: Section 2 contains the details of the neural network architecture based on the geometrical properties of CMS detector image. Section 3 consists of two parts. In the first part, the theoretical description of the RPV SUSY process, the baseline selection applied to select events corresponding to final state of interest are presented. In the second part, overall processes applied for using the image data used as the training inputs are described along with a summary of the number of training inputs. Section 4 presents details on the distributed training using the Nurion HPC system. Section 5 elucidates the results of this study, including a comparison between the classification performance of CNN and cut-based method, as well as the scalability of the CNN model in the Nurion HPC system. Section 6 summarizes the main findings of the study.


\section{CNN architecture with circular symmetry}
In general, cylindrical detectors are used in HEP experiments, and therefore, a cylindrical coordinate system is used to define the geometric position of the detector. The CMS detector also follows this system. The z component indicates the beam direction in the LHC ring, the radical component in the transversal plane of the beam direction is denoted by r, the azimuthal angle is represented by $\phi$, and the pseudorapidity is denoted by $\eta$, which is the Lorentz invariant polar angle defined as $\eta = -\ln[\tan(\frac{\theta}{2})]$. Because the detector image has a circular symmetry according to the $\phi$ direction, the symmetry should be considered while building the neural network structure, instead of building a typical two-dimensional (2D) image classification model. Therefore, we focus on the padding method that works well with the circular symmetry of detector images.
  
\begin{figure}[ht]
   \begin{adjustbox}{varwidth=\textwidth,margin=0 {\abovecaptionskip} 0 0, frame=0.1pt }
   \centering\includegraphics[scale=0.45]{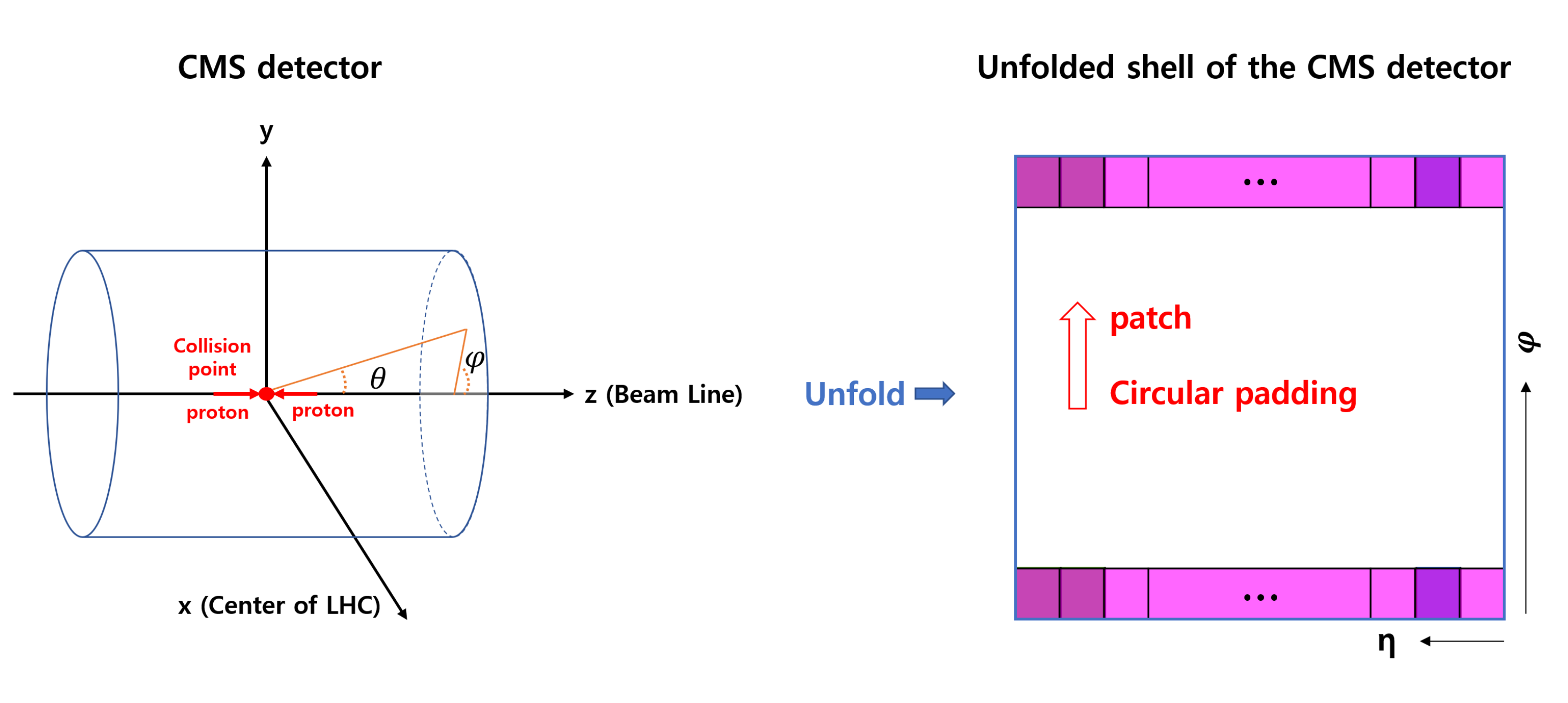}
   \bfseries
   \caption{Illustration of the ``circular padding'' scheme in detector image which has symmetry at $\phi$ direction. The ``zero padding'' is applied on the $\eta$ direction which is defined as $\eta= -\ln[\tan(\theta/2)]$. In the $\phi$ direction, however, the bottom row of pixel is pasted as the top row of pixel in order to prevent the loss of information at the $\eta$ direction as a result of ``zero padding''.}
       \label{fig:CircPad}
   \end{adjustbox}

\end{figure}

The CNN models incorporate the ``zero padding'' process to maintain the size of the feature map by attaching zeros to outline of the image. However, this can lead to loss of information along $\eta$ direction, because of the $\phi$ symmetry. Therefore, we adopt a new padding technique, namely the ``circular padding'', wherein we still use the ``zero padding'' in the $\eta$ direction, however, in the $\phi$ direction, the bottom row of pixels is pasted as the top row of pixels. The circular-padding scheme is illustrated in Figure \ref{fig:CircPad}
  
\begin{figure}[ht]
  \begin{adjustbox}{varwidth=\textwidth,margin=0 {\abovecaptionskip} 0 0, frame=0.1pt }
  \centering\includegraphics[scale=0.5]{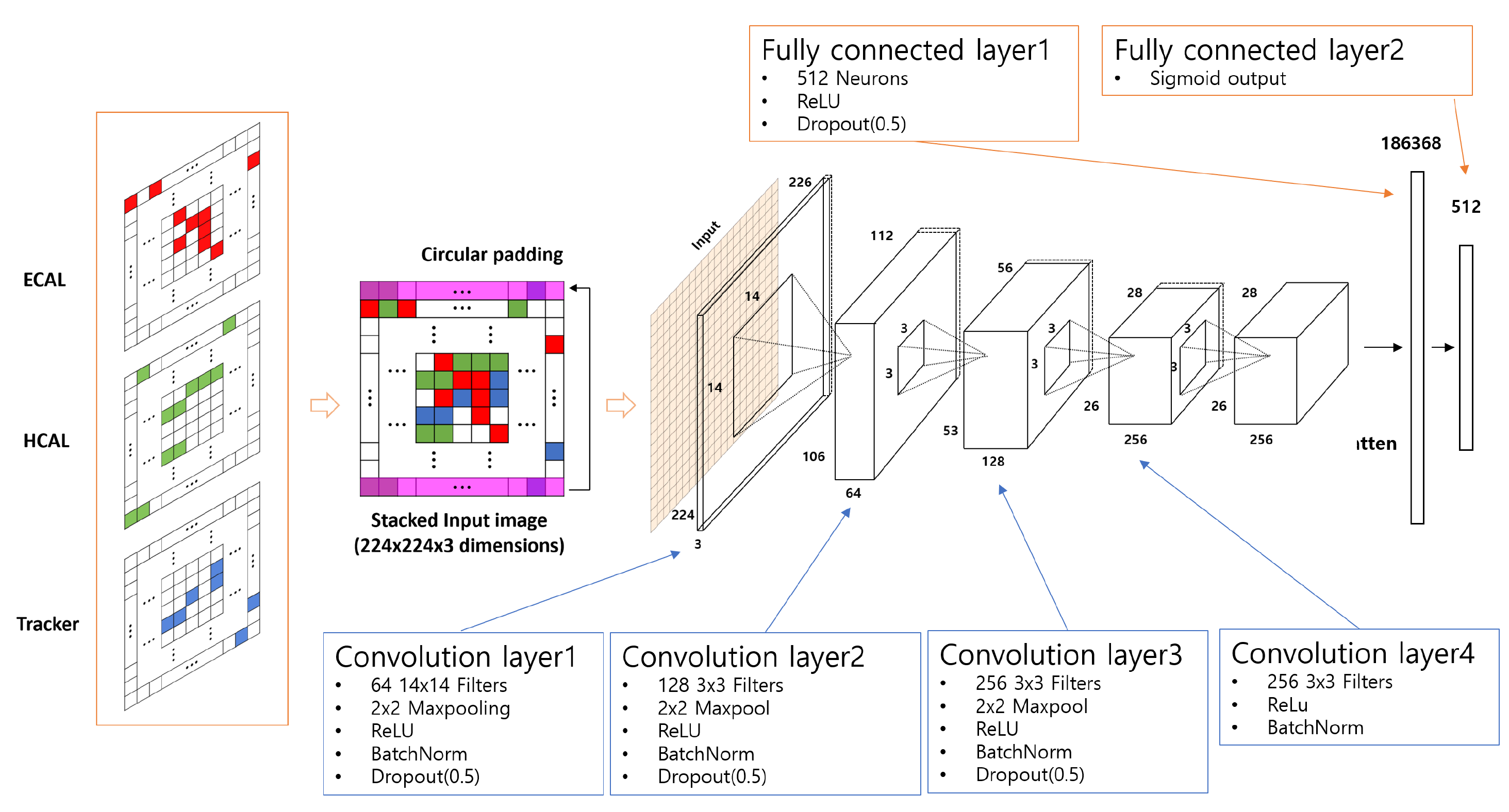}
  \bfseries
  \caption{Illustration of the CNN architecture. Before the training step, image data are prepared by merging three sub-detector images: tracker, electromagnetic calorimeter (ECAL), and hadronic calorimeter (HCAL). The ECAL and HCAL images are weighted by energy measured in the detector, tracker image is weighted by transverse momentum ($\mathrm{p_{T}}$), and finally these three images are merged as an image with 224x224x3 dimension. The CNN model is trained to do binary classification: RPV SUSY signal and QCD multi-jet background. There are four convolution layers and two fully connected layers with Max pooling, ReLu activation function, batch normalization and dropout in order. The ``circular padding'' is applied on all convolution layers. Two fully connected layers are located at the end of the architecture to flatten the information and do a binary classification using the sigmoid activation function. This architecture is used to train the $224\times224$ images the final result of this paper}
    \label{fig:CNNarc}
  \end{adjustbox}

\end{figure}

The CNN architecture used to classify the $224\times224$ images is depicted in Figure \ref{fig:CNNarc}, indicating four convolution layers and two fully connected layers. Max pooling \cite{maxpooling}, the ReLu activation function \cite{ReLu}, batch normalization \cite{Batchnormal} and dropout \cite{srivastava2014dropout} are used in the first three convolution layers sequentially. Although we also test another sequence such as ReLU-batch normalization-Max pooling-dropout, the current order shows the best signal efficiency. The fourth convolution layer only contains ReLu and batch normalization. In the convolution operation, $3\times3$ and $14\times14$ filters are used, whereas the $14\times14$ filter is used only in the first layer to obtain a large amount of information before compressing the information in the pooling steps. The filters move one stride in the convolution layers and two strides in the Max pooling steps, where a $2\times2$ filter is used. We randomly turn off 50\% of the neurons in the dropout layer at the training stage, whereas use all the neurons at the validation and test stages. 

When we train the $64\times64$ images, the same CNN architecture used in training the $224\times224$ images is employed, however, the ``circular padding'' and $14\times14$ filters are not used. In the fully connected layers, 512 neurons with batch normalization and dropout are used. Moreover, the sigmoid activation function is used in the output layer for binary classification, and the cross-entropy loss is used with the \textsc{adam} optimizer \cite{adam}. We optimize the learning rate based on 64 nodes, with a scan range from 0.001 to 0.01 at intervals of 0.001, and optimized learning rate of 0.02 (0.008) for the $224\times224$ ($64\times64$) images.

\section{Benchmark model and datasets} 

The SUSY is motivated to solve problems such as dark matter and mass hierarchies that cannot be explained using the SM. This model introduces new symmetry between bosons and fermions  \cite{SUSY_REF1,SUSY_REF2,SUSY_REF3,SUSY_REF4,SUSY_REF5,SUSY_REF6,SUSY_REF7,SUSY_REF8}, and predicts new particles corresponding to the SM fermions and bosons. The \textit{R} parity is a new quantum number that enforces the conservation of lepton and baryon numbers in the SUSY model through global symmetry \cite{Rparity}. This parameter is defined as $R_{p} = (-1)^{2s + 3(B-L)}$, where $s$ is the spin, $B$ is the baryon number and $L$ is the lepton number. If the \textit{R}-parity is conserved, then lightest SUSY particles (LSPs) become stable, and this leads to a signature of large amount of missing transverse energy because the LSPs are undetectable. In contrast, if the \textit{R}-parity is violated (RPV), the LSPs become unstable and decay to known SM particles. Therefore, the absence of missing transverse energy from undetectable LSPs is the experimental advantage of the RPV SUSY model.
  
In this study, we focus on a RPV gluino pair production as signal process. In this process, each gluino decays as follows: $\tilde{g} \rightarrow \tilde{t} \rightarrow tbs$ . It is assumed that the mass of the gluino ($\tilde{g}$) equals 1.4 TeV and the stop squark ($\tilde{t}$) is much heavier than the gluino. Therefore, the stop squark is off-shell in this decay and gluino is the LSP. The top quark decays to a b-jet which is the jet containing a B hadron and two hadron jets, resulting in 10 jets in the final state. As mentioned earlier, the significant number of SM backgrounds make it difficult to search the SUSY process. The major SM background in this analysis is QCD multi-jet process which also has almost 10 jets in the final state. 
  
The proton-proton collision events with a center-of-mass energy of 13 TeV in the LHC ring are simulated using the \textsc{pythia 8} \cite{sjostrand2015introduction} program.
The detector simulation assuming CMS detector is performed using the $\textsc{delphes 3.4}$ program \cite{de2014delphes} with the default ``CMS detector configuration'', and $\num{330000}$ RPV SUSY events and \num{20000000} QCD multi-jet events are generated. Furthermore, an average of 32 interactions per proton bunch crossing (pile-up) is considered.

\begin{table}[h]
\caption{\label{tabone}Summary of dataset: The QCD multi-jet samples are generated as different $\mathrm{H_{T}}$ ranges. The number of selected events is the remaining events after passing the baseline selection. These events are used as inputs of DNN model. There are roughly \num{450000} training images, \num{150000} validation images, and \num{150000} testing images. In addition, background samples are normalized and scaled to the weighted sum of backgrounds equal to the number of signal selected events in the training step. The QCD multi-jet sample with $\mathrm{H_{T}} <$ 1000 GeV are all excluded after the baseline selection.} 
\begin{adjustbox}{width=\textwidth}
\lineup
\begin{tabular}{@{}*{4}{l}}
\br                              
dataset& Cross section [pb]& Number of generated events & Number of selected events\cr 
\mr
RPV                         &0.02530    &\num{330599}    &\num{294762}\cr
QCD ($\mathrm{H_{T}}$ 1000-1500 GeV)      &\num{1207}             &\num{15466225} &\num{37091}\cr 
QCD ($\mathrm{H_{T}}$ 1500-2000 GeV)      &119.9          &\num{3368613}  &\num{137805}\cr 
QCD ($\mathrm{H_{T}}$ 2000-Inf GeV)       &25.24        &\num{3250016}  &\num{280279}\cr 
\br
\end{tabular}
\end{adjustbox}
   \label{tab:dataset}
\end{table}
\normalsize

Table \ref{tab:dataset} shows the summary of the dataset. The QCD multi-jet processes are generated using different ranges of $\mathrm{H_{T}}$, which is defined as the scalar sum of the transverse momentum ($\mathrm{p_{T}}$) of all the jets in an event, to gain statistics in the high $\mathrm{p_{T}}$ region. The baseline selection is primarily applied to select the event of interest, and the QCD multi-jet samples with $\mathrm{H_{T}} <$ 1000 GeV are rejected after the baseline selection. The events passing the baseline selection denoted as the ``selected event'' in Table 1 are used as the CNN inputs. There are approximately \num{750000} images in each of two categories: RPV SUSY signal and QCD multi-jet background. The dataset is divided by \num{450000} training images, \num{150000} validation images, and \num{150000} testing images. The physical variables used in the baseline selection are $\mathrm{p_{T}}$, $\mathrm{H_{T}}$, and $\eta$, mass sum of the fat jets (with larger cone-size ($\Delta R$=1.2) \cite{jetalgoreview} compared to those of the jets ($\Delta R$=0.4)) \cite{jetalgoreview}, and the number of jets and b-jets. 
  
The search for RPV SUSY process for the same final state was studied in CMS experiment at 13 TeV \cite{cms2016search} using the cut-based method. We refer all the cut criteria and physical variables from \cite{cms2016search}. Finally, the signal efficiency obtained from the CNN method is compared with that obtained from the cut-based method \cite{cms2016search}. To obtain results of the cut-based method, the signal-region (physics selection) selected in \cite{cms2016search} is used. In this method, the events passing the physics selection are classified as the signals. All cut-criteria are summarized in Table \ref{tab:Selections}.
  
\begin{table}[h]
\caption{\label{jlab1}Summary of selections. The baseline selection is primarily applied to select event of interest. The physics selection is applied to select signal events in cut-based method. The jets are defined as $\Delta R$=0.4 and $\mathrm{p_{T}} > 30$ GeV and $|\eta| \leq 2.4$. The fat jets are defined as $\Delta R$=1.2 and $\mathrm{p_{T}} >$ 30 GeV}
\centering
\begin{tabular}{@{}l|l}
\br
Baseline selection      & Physics selection       \\ 
\mr
At least 4 jets         & At least 8 jets         \\
Mass sum of fat jet $> 500$ GeV  & Mass sum of fat jet $> 800$ GeV  \\
$\mathrm{H_{T}}$ \textgreater 1.5 TeV & $\mathrm{H_{T}}$ \textgreater 1.5 TeV \\
At least 1 B-jet        & At least 3 B-jets       \\ 
\br
\end{tabular}\\
\label{tab:Selections}
\end{table}
\normalsize

\begin{figure}[h]
\begin{adjustbox}{varwidth=\textwidth,margin=0 {\abovecaptionskip} 0 0, frame=0.1pt }
\centering\includegraphics[scale=0.65]{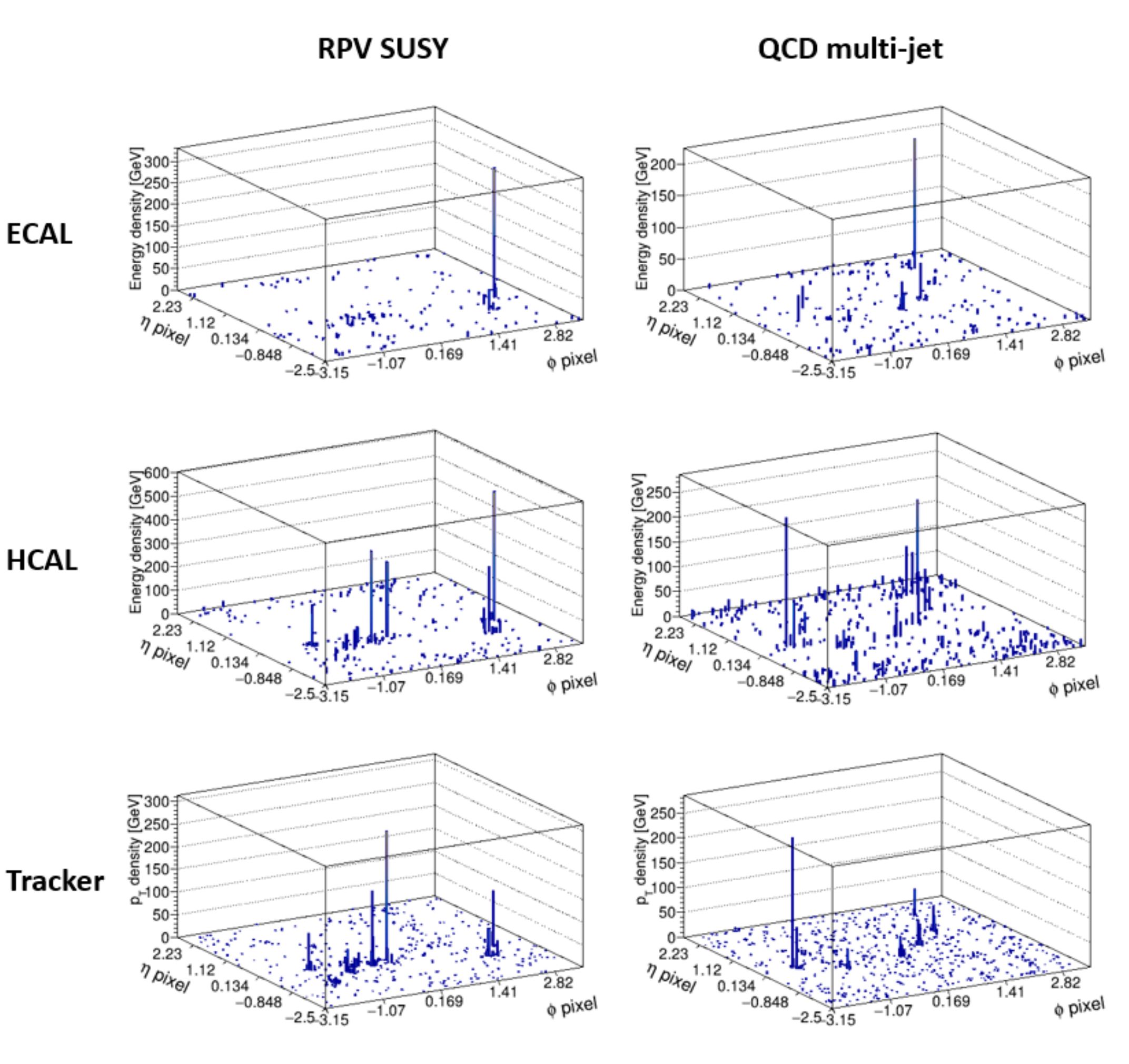}
\caption{Single RPV SUSY event and single QCD multi-jet event in all sub-detectors are visualized as 3D images with 224x224 pixel. Each image are 3D-rendered at z direction which means the energy density ($\mathrm{p_{T}}$ density) of ECAL and HCAL (tracker). In practice, the image used for training is represented by energy or $\mathrm{p_{T}}$ as pixel intensity. In addition, HCAL, ECAL, and tracker are represented as one image of three channels of color: red, green, and blue. The trained CNN model can discriminate RPV SUSY signal from QCD multi-jet background using this single image data.}
\label{fig:Visualize}
\end{adjustbox}
\end{figure}

The information of final-state particles (10 jets) is recorded in the sub-detectors of the CMS detector, i.e. the tracker, electro-magnetic calorimeter (ECAL), and hadronic calorimeter (HCAL). The data from these sub-detectors can be interpreted as 2-D images corresponding to an azimuthal angle $|\phi|$ $<$ 3.15 and a pseudorapidity $|\eta|$ $<$ 2.5. 
The 2-D histograms of $|\phi|$ and $|\eta|$ illustrated in Figure \ref{fig:Visualize} show the images of the single events with 3-D rendering results in the z-axis (that is, energy density ($\mathrm{p_{T}}$ density) for ECAL and HCAL (tracker)). This technique is similar to the approach reported in \cite{bhimji2017deep}. However, we applied further improvements to increase the signal efficiency. The details of the process for developing the image data are provided below.  
\begin{enumerate}
  \item Formation of images of the particles deposited in the detectors: First, 2-D histograms using the number of events in $\eta - \phi$ plane are produced. The bins are set as 224 (or 64), which are then converted to images with resolutions of $224\times224$ and ($64\times64$) pixels.  
  \item Weighting image by energy or $\mathrm{p_{T}}$: The HCAL and ECAL images are weighted by energy, whereas the tracker images are weighted by $\mathrm{p_{T}}$. Therefore, pixel intensity of the HCAL and ECAL images (tracker images) represent energy density ($\mathrm{p_{T}}$ density).
  \item Three channel merging: The tracker, HCAL and ECAL images are treated as three colored layers: red, green and blue, which are merged to form a single image with three channels. Therefore, the dimensions of an image are $224\times224\times3$ ($64\times64\times3$).
  \item Normalization and rescaling: In HEP, the number of expected events can be obtained using the following equation:
\begin{eqnarray}
\fl \mathrm{Number\, of\, expected\, events} = \sigma \times L \times \mathrm{\frac{Number \, of \, selected \, events}{Number \, of \, generated \, events}} \label{eq1}
\end{eqnarray}

where $\sigma$ is the cross-section of the physics process (calculated in \textsc{PYTHIA 8} generator); this is proportional to the probability of generating the process, and $L$ represents the integrated luminosity, which is proportional to the size of data.
  
As shown in the Table \ref{tab:dataset}, the QCD multi-jet backgrounds are generated with three different $\mathrm{H_{T}}$ ranges, and therefore, have three different cross-sections even though they are same processes. To consider this difference, normalization is applied using a weight factor expressed as ``$w =$ $\sigma$ $/$ Number of generated events'', based on equation (\ref{eq1}). After the normalization, the background events are rescaled to match the weighted sum of background events, which is equal to the number of selected signal events in Table \ref{tab:dataset}. This rescaling is applied to balance the number of input signal with that of the input background. 
\end{enumerate}

(\lowercase\expandafter{\romannumeral3}) and (\lowercase\expandafter{\romannumeral4}) are implemented in the training stage on the fly. The resulting pre-processed images are divided by training, validation, and test dataset at the ratio of 6:2:2 to prevent overfitting and used as the input in the large scale training described in the next section.

\section{Large-scale training}
The distributed training is performed in the Nurion HPC system using the Horovod framework \cite{sergeev2018horovod}. The training model is copied to all the worker nodes, and the data are divided and distributed to all the nodes. When the training stage begins, the gradient of each node is calculated for each batch, and when the gradient calculations of all the nodes are completed, the model is updated with the average of all gradients (``synchronous distributed training'') \cite{goyal2017accurate}. The nodes are linked in a ring-shape and each node sends its gradients to its successive neighbor on the ring (``ring all-reduce architecture'') \cite{sergeev2018horovod}. The advantage of this setting is that it scales up as well as speeds up proportional to the amount of available data \cite{paine2013gpu}.

\begin{table}[h]
\caption{\label{jlab1}Batch size and the number of nodes in ``strong scaling'': overall batch sizes equal to 32K}
\centering
\footnotesize
\begin{tabular}{@{}llc}
\hline
Number of nodes      & Batch size & Effective batch size       \\ 
\hline
64   & 512  \\
128  & 256  \\
256  &  128 & \num{32768} \\
512  & 64   \\ 
1024  & 32   \\ 
\hline
\end{tabular}\\
   \label{tab:Nodes}
\end{table}
\normalsize

Because data parallelism and ``synchronous distributed training'' are used, the overall batch size is equal to batch size times the number of nodes. Therefore, the training model is scaled up by conserving the overall batch size ($\mathrm{=32K}$), i.e. ``strong scale''. The number of nodes and batch sizes used in the ``strong scaling'' are listed in Table \ref{tab:Nodes}. Further, the results obtained with the scaled-up and fixed batch size with 8 batches (``weak scaling'') are compared with those obtained with the ``strong scaling''. The main results obtained ``strong scale'' are shown in the next section.

\section{Results}
\begin{figure}[h]
\begin{adjustbox}{varwidth=\textwidth,margin=0 {\abovecaptionskip} 0 0, frame=0.1pt }
\begin{subfigure}{.5\textwidth}
  \centering
  \includegraphics[width=.8\linewidth]{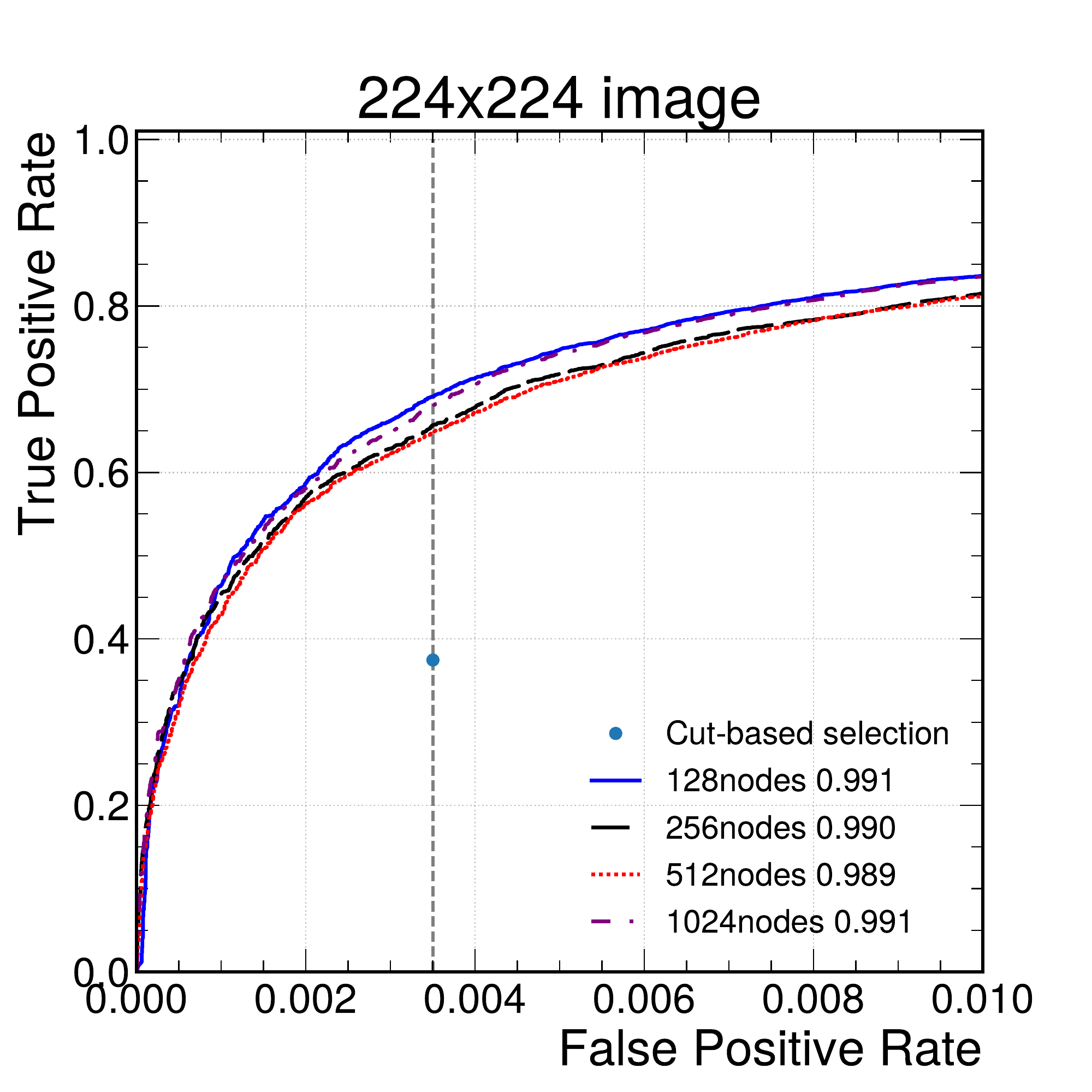}
  \label{fig:sfig1}
\end{subfigure}%
\begin{subfigure}{.5\textwidth}
  \centering
  \includegraphics[width=.8\linewidth]{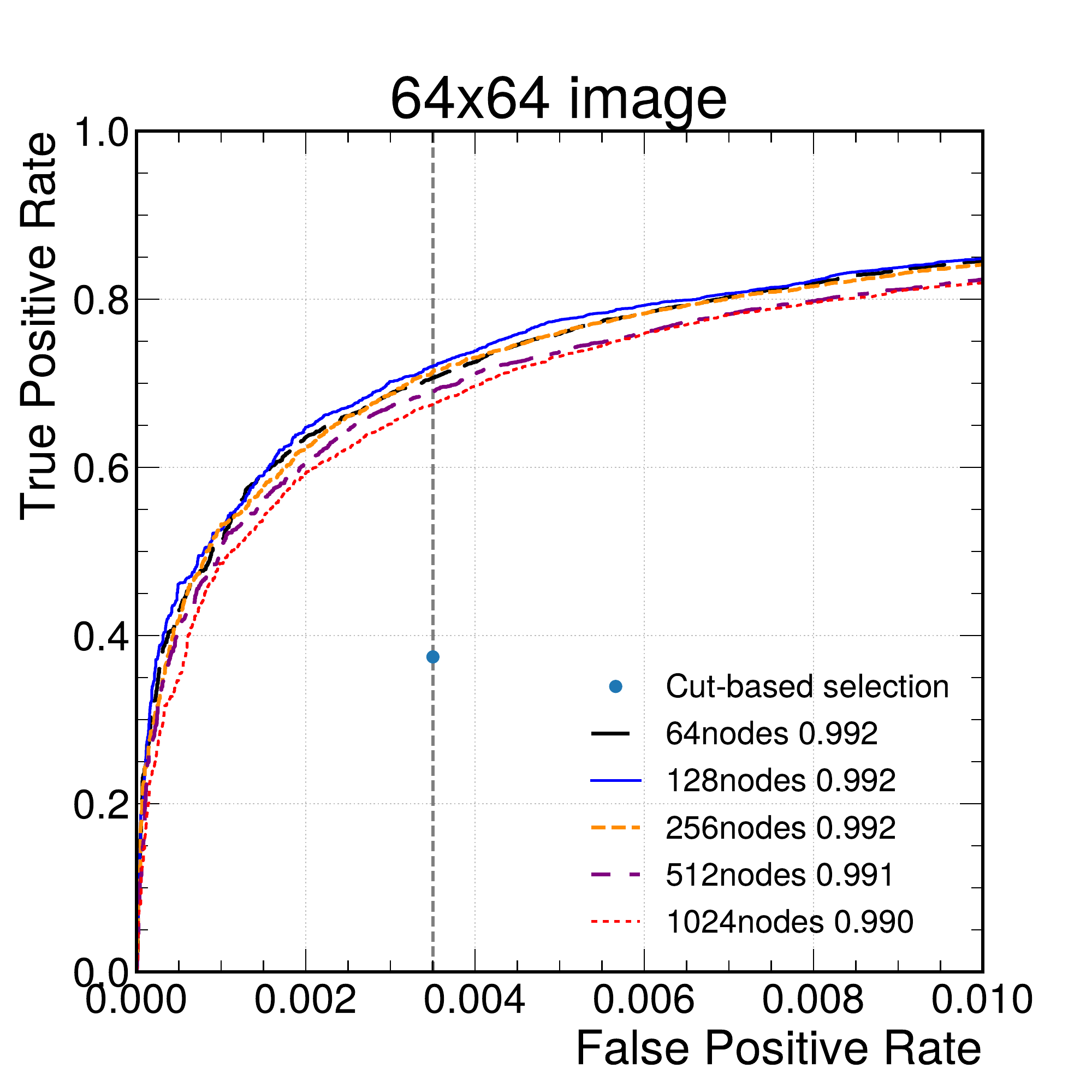}
  \label{fig:sfig2}
\end{subfigure}
\caption{ROC curve of training results from $224\times224$ (left) and $64\times64$ (right) images. ROC curves of CNN results with a different number of nodes are represented as different types of lines. The TPR point in the given FPR from the cut-based method is represented as a blue point. Each AUC value according to the number of nodes is shown in the legend. The results of the CNN method with multi-nodes show higher signal efficiency (TPR) than that of the cut-based method in both $224\times224$ and $64\times64$ cases. Training results from both $64\times64$ and $224\times224$ images show similar results. The training using 128 nodes shows the highest AUC for both images. Also, each node shows a similar ROC curve, which means that performances of multi-node are stable}
\label{fig:ROC_curve}
\end{adjustbox}
\end{figure}

\begin{figure}[ht]
\begin{adjustbox}{varwidth=\textwidth,margin=0 {\abovecaptionskip} 0 0, frame=0.1pt }
\begin{subfigure}{.5\textwidth}
  \centering
  \includegraphics[width=\linewidth]{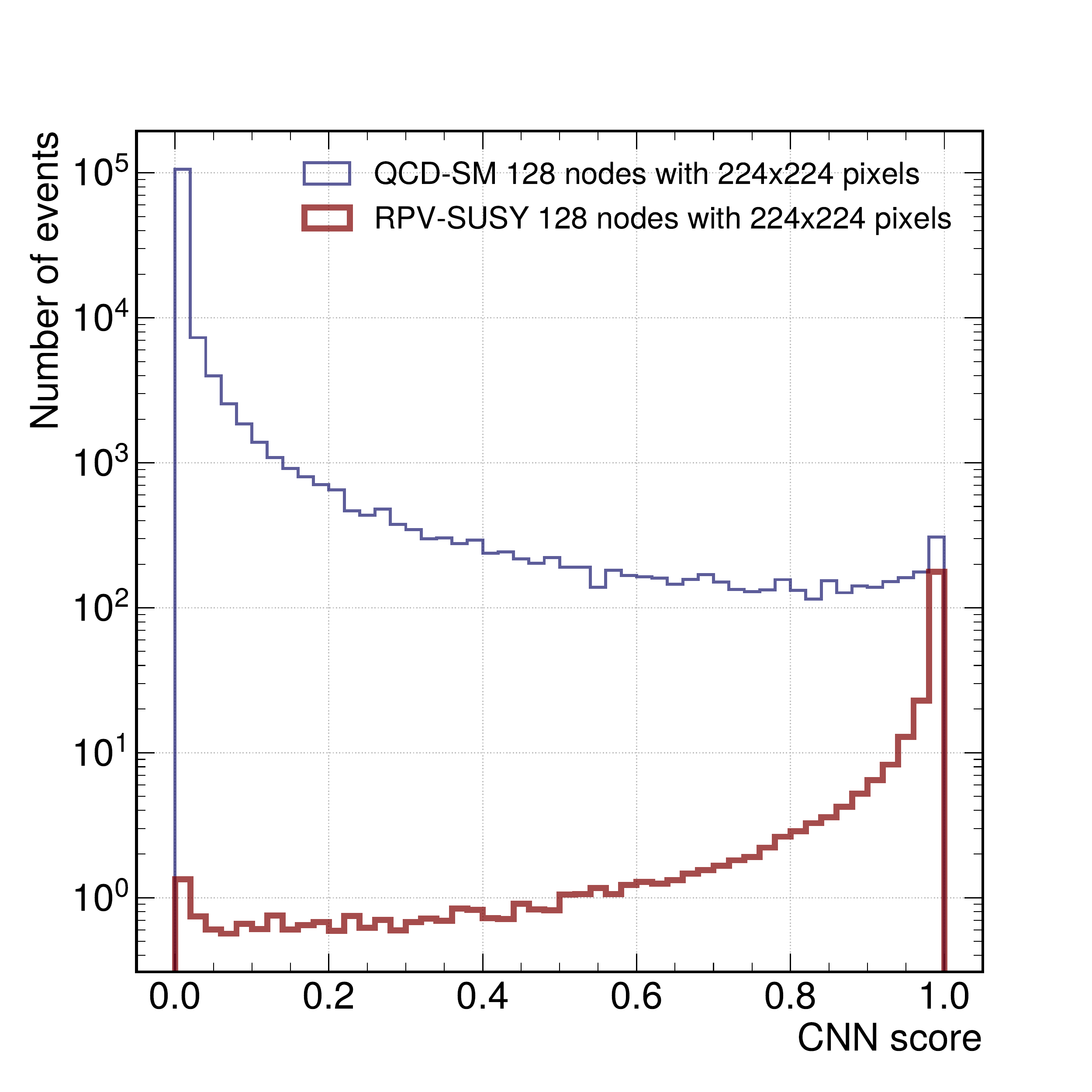}
  \label{fig:sfig1}
\end{subfigure}%
\begin{subfigure}{.5\textwidth}
  \centering
  \includegraphics[width=\linewidth]{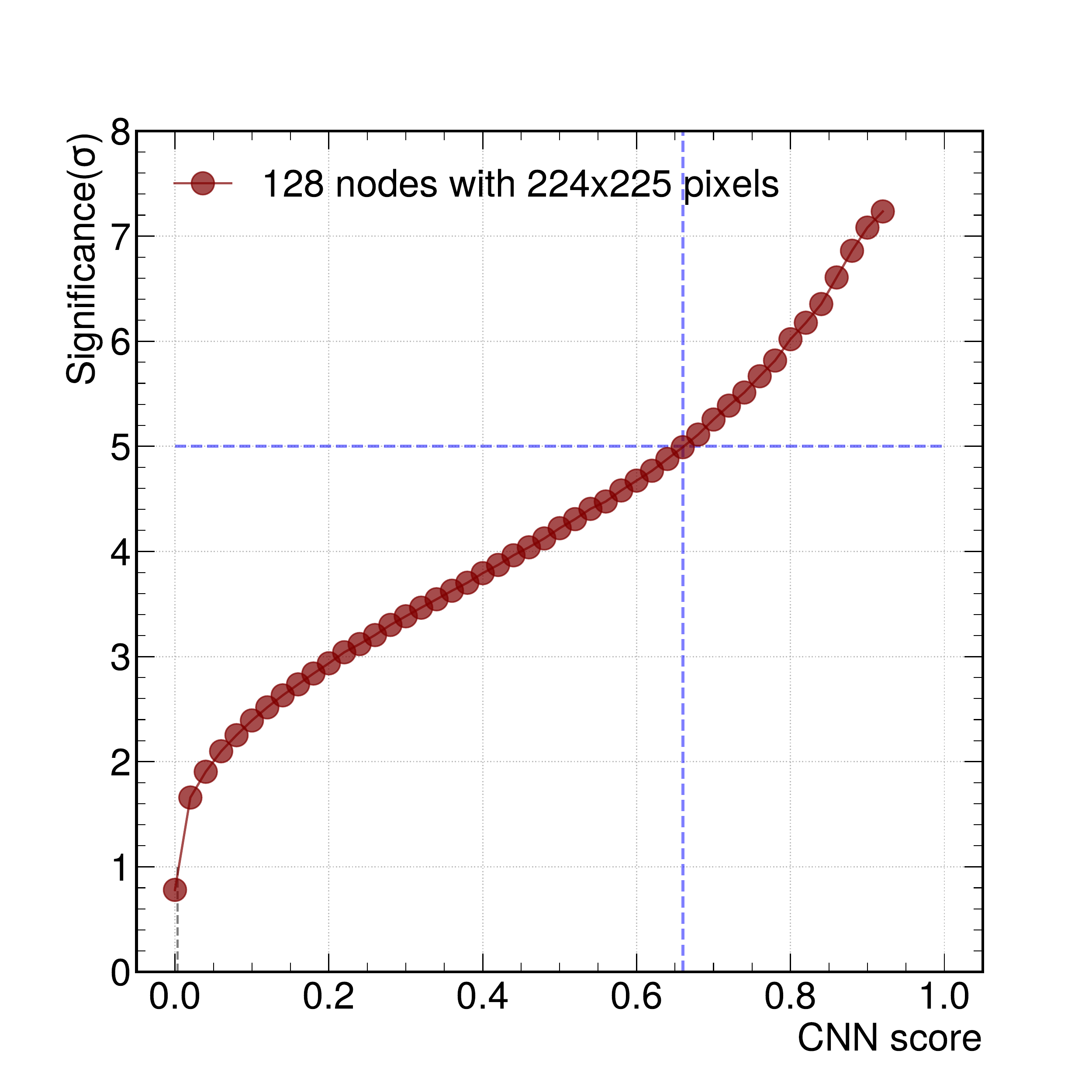}
  \label{fig:sfig1}
\end{subfigure}%
\caption{Distributions of CNN score and corresponding expected significance as a function of the CNN scores, calculated from the results of $224\times224$ images with 128 nodes. The left panel shows the normalized CNN score using equation (\ref{eq1}) whose area under the histogram equals the number of expected events. The right panel is the expected significance calculated from the hypothesis testing using the expected number of events from the left histogram. The discovery level (5$\sigma$) is represented as a dotted line. This experiment reaches the discovery significance at a CNN score of 0.68 with 258 signal events and 2476 background events.}
\label{fig:Score}
\end{adjustbox}
\end{figure}

The performance of the training model is evaluated based on the true positive rate (TPR) and false positive rate (FPR), which mean the signal efficiency and 1-background rejection, respectively. These values are visualized using the receiving operator characteristic curve (ROC curve) in Figure \ref{fig:ROC_curve}. The ROC curve is constructed using a CNN score plot as shown in the left panel of Figure \ref{fig:Score}. First, we plot the CNN score based on the predicted values from test dataset. Next, we set the threshold on the CNN score plot and treat the right-hand side of the threshold as a signal, and the left-hand side of the threshold as the background. Then, we integrate the numbers in each region. Finally, we continuously calculate the TPR and FPR while scanning the threshold from 0 to 1 and construct the ROC curve. The area under the ROC curve (AUC), whose ideal value is 1, indicates how good is the signal-background separation. We obtain an average AUC of 0.9903 (0.9914) for the $224\times224$ ($64\times64$) pixel images. The training using 128 nodes show the highest AUC for both the $224\times224$ and $64\times64$ pixel images. Furthermore, each node shows a similar ROC curve, indicating that the performances of multi-nodes are stable.

We compare the signal efficiency of the CNN and cut-based method at the same background rejection point (FPR=0.0035) by considering the training result obtained using 128 nodes, because it shows the best performance. The signal efficiency (TPR) using the CNN method is 1.845 (1.803) times higher than that obtained using the cut-based method for the $224\times224$ ($64\times64$) pixel images.


The another method to check the model performance is the expected significance. The expected significance verify not only the training performance but also physical meaning, because it shows the possibility of discovery if signal process. We calculate the expected significance using the following steps.

\begin{enumerate}
   \item Normalize the CNN score plot (left panel in Figure \ref{fig:Score}) such that the area under the histogram becomes equal to the number of expected events calculated using equation (\ref{eq1}). 
   \item Set the threshold on the CNN score plot and calculate the number of expected events in the signal-region.  
   \item Perform a hypothesis testing using Poisson distribution, i.e. null hypothesis: background only, and alternative hypothesis: signal+background.  
   \item Calculate the p-value and convert it to the expected significance.
   \item Scan the threshold from 0 to 1, and draw the CNN score vs the significance plot.
\end{enumerate}

The expected significance according to the CNN score is shown in the right panel of Figure \ref{fig:Score}. We compare the expected significance of the cut-based method with that of the CNN method at the same background rejection point (FPR=0.003503). The CNN method raises the expected significance near 2$\sigma$, which is 1.2 times higher than that of the cut-based method. The expected significance and related statistical values are listed in Table \ref{tab:Significance}.

\begin{table}[ht]
\caption{\label{jlab1}Comparison of expected significance between CNN (128 nodes) and cut-based method within same background rejection point (FPR=0.003503). The expected significance of the CNN method is 1.2 times higher than that of the cut-based method.}
\centering
\footnotesize
\begin{adjustbox}{width=\linewidth}
\begin{tabular}{@{}llll}
\hline
                                       & Cut-based & CNN ($64\times64$) & CNN ($224\times224$) \\ 
\hline
\multicolumn{1}{l}{Significance}     & 1.55    & 1.79     & 1.80      \\ 
\multicolumn{1}{l}{P-value}          & 0.0608 & 0.0366   & 0.0356     \\ 
\multicolumn{1}{l}{Number of expected RPV SUSY events}     & \num{553}       & \num{639}         & \num{664}           \\ 
\multicolumn{1}{l}{Number of expected QCD-multijet events} & \num{126771}   & \num{126686}     & \num{126681}       \\ 
\hline
\end{tabular}
\end{adjustbox}
   \label{tab:Significance}
\end{table}

The main purpose of calculating the significance is to investigate whether this process can be discovered or not. Therefore, we investigate the number of expected events in the discovery level (5$\sigma$). As shown in Figure \ref{fig:Score}, we find an optimal point of CNN score that reaches the 5$\sigma$ level, and calculate the number of expected events of signal and background. There are 258 (259) signals and 2476 (2473) backgrounds, for $224\times224$ and $64\times64$ pixel images.
  
The $224\times224$ pixel image, owing to its high resolution, is used to minimize the information loss, thereby improving the training performance. However, the classification performance using the $224\times224$ pixel images does not show any significant enhancement. The AUC of the $224\times224$ images is slightly lower than that of the $64\times64$ images, and the expected significance of the best-performing node, obtained from the $224\times224$ image, is slightly higher than that of obtained from the $64\times64$ image. However, we demonstrate the scalability using the $224\times224$ image.

\begin{figure}[h]
\begin{adjustbox}{varwidth=\textwidth,margin=0 {\abovecaptionskip} 0 0, frame=0.1pt }
\begin{subfigure}{.5\textwidth}
  \centering
  \includegraphics[width=\linewidth]{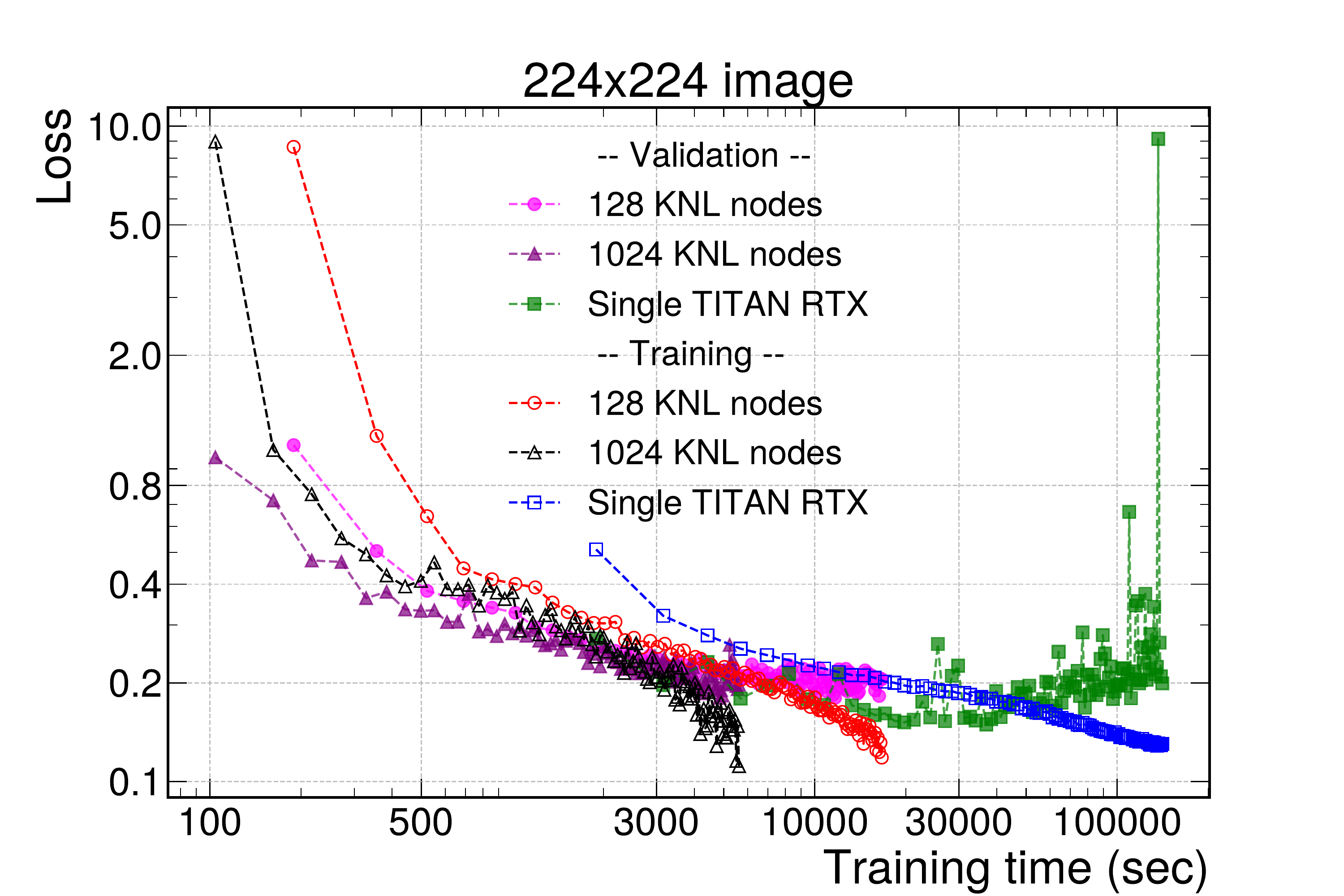}
  \label{fig:sfig1}
\end{subfigure}%
\begin{subfigure}{.5\textwidth}
  \centering
  \includegraphics[width=\linewidth]{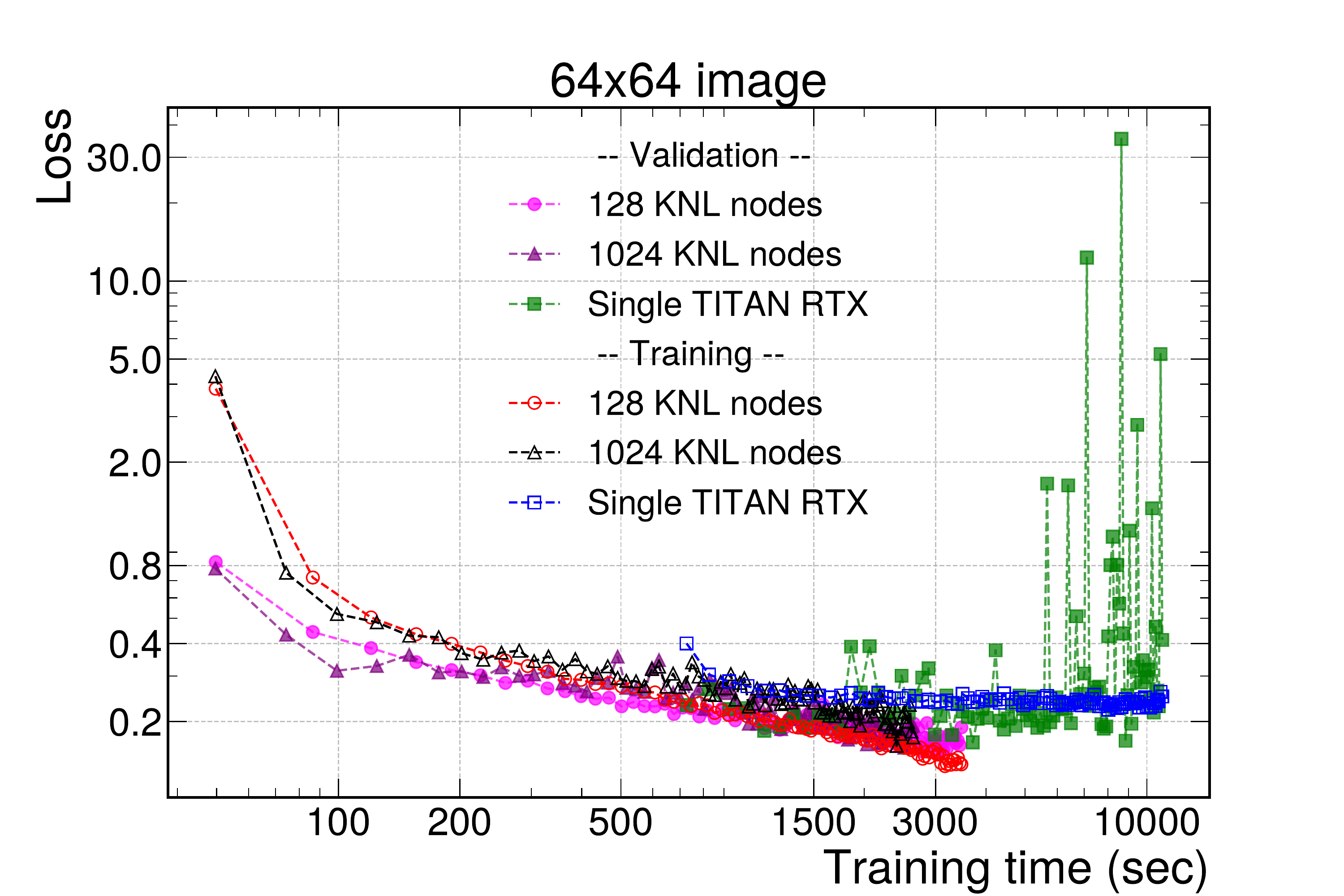}
  \label{fig:sfig1}
\end{subfigure}%

\caption{Training time per loss using $224\times224$ images. The X-axis is the training time and the Y-axis is the loss. The training curves of 128 and 1024 number of the KNL nodes and a Single GPU machine are compared using the same training and validation dataset. This experiment proceeded with loaded all training data in the main memory of the Nurion HPC system before the beginning of the training. In order to check the time with an increasing number of nodes, data loading time is excluded. On the other hand, only the batch with show the lowest validation loss are considered in updating the CNN model to deal with overfitting.}
\label{fig:LearningCurveGPU}
\end{adjustbox}
\end{figure}

Next, the training is evaluated to examine the scalability of the CNN model. All the scalability tests are performed by loading all the training data in the main memory of the Nurion HPC system, before the beginning of the training. To evaluate variations in the training time with increasing number of nodes, the data loading time is excluded. Figure \ref{fig:LearningCurveGPU} shows a learning curve that indicates the loss according to the training time. The learning curve of a single GPU (TITAN RTX) is compared with the results obtained from multi-nodes. Notably, the single GPU shows the lowest loss but takes the longest training time. In contrast, the multi-nodes show similar or slightly lower performance than that of a single GPU. However, the training time of the multi-nodes is significantly reduced compared to that of a single GPU. The training with 1024 KNL nodes is 25 times (4 times) faster than that of a single TITAN RTX node using $224\times224$ ($64\times64$) pixel images.

\begin{figure}[ht]
\begin{adjustbox}{varwidth=\textwidth,margin=0 {\abovecaptionskip} 0 0, frame=0.1pt }
\begin{subfigure}{.5\textwidth}
  \centering
  \includegraphics[width=.9\linewidth]{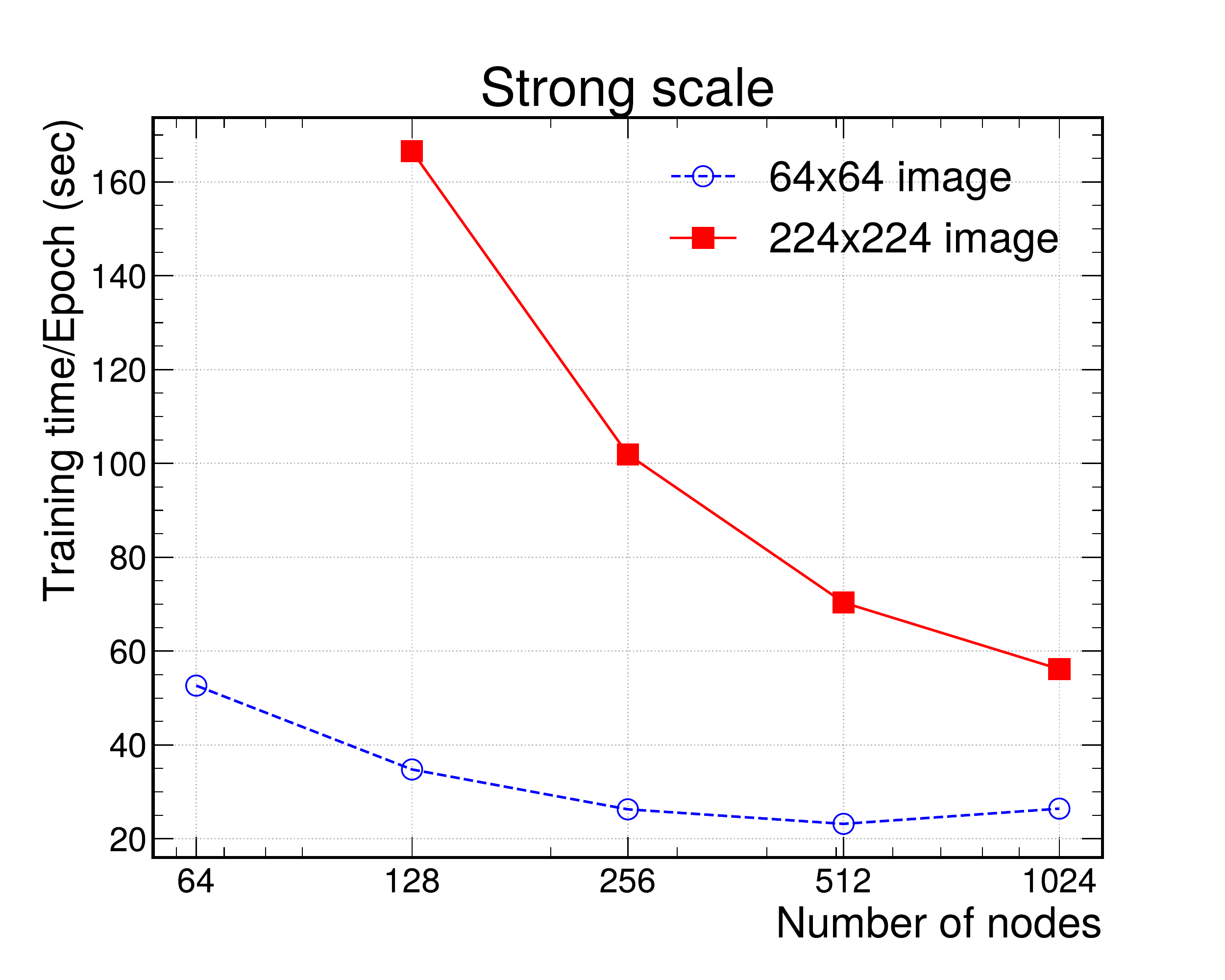}
  \label{fig:sfig1}
\end{subfigure}%
\begin{subfigure}{.5\textwidth}
  \centering
  \includegraphics[width=.9\linewidth]{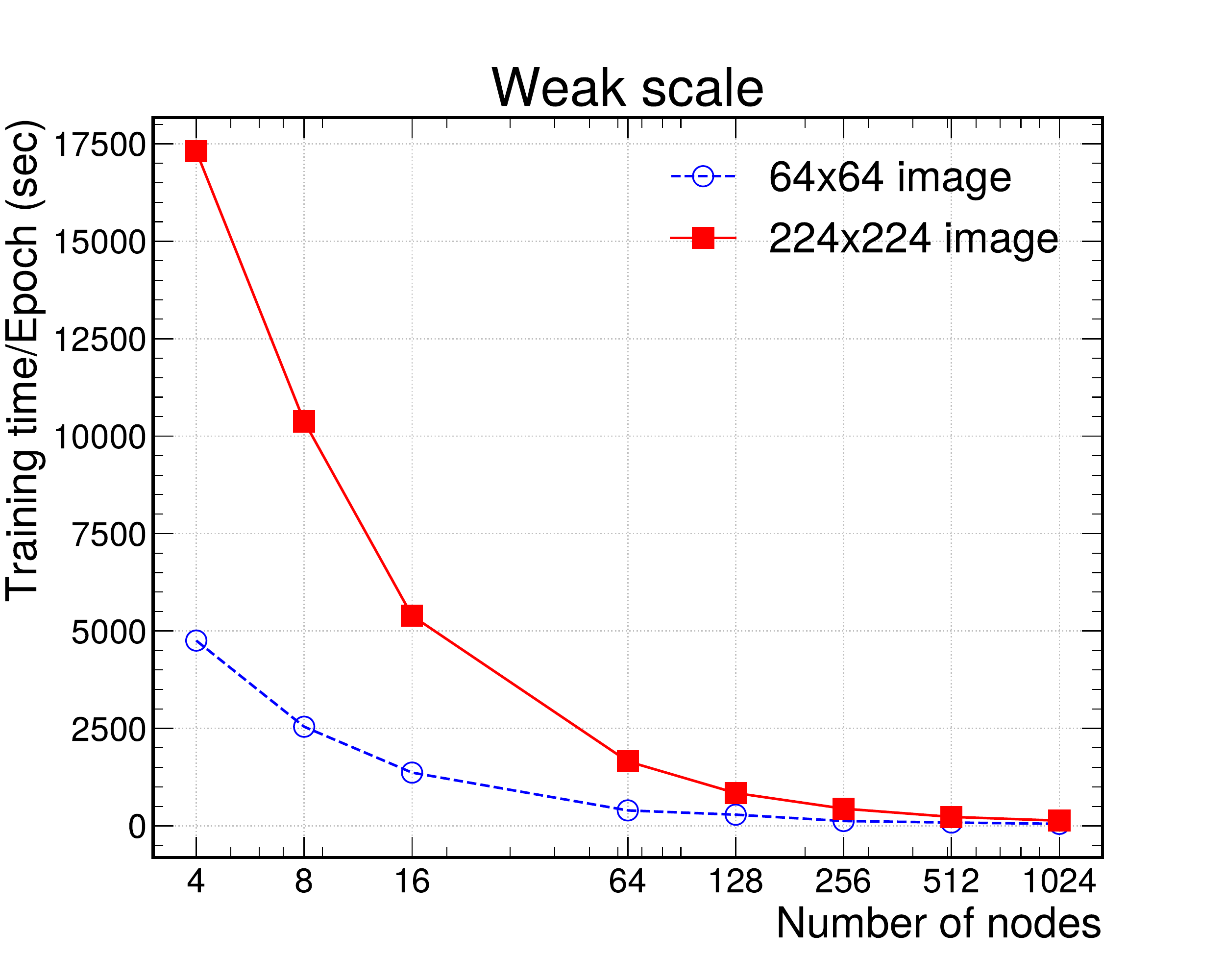}
  \label{fig:sfig2}
\end{subfigure}
\begin{subfigure}{.5\textwidth}
  \centering
  \includegraphics[width=.9\linewidth]{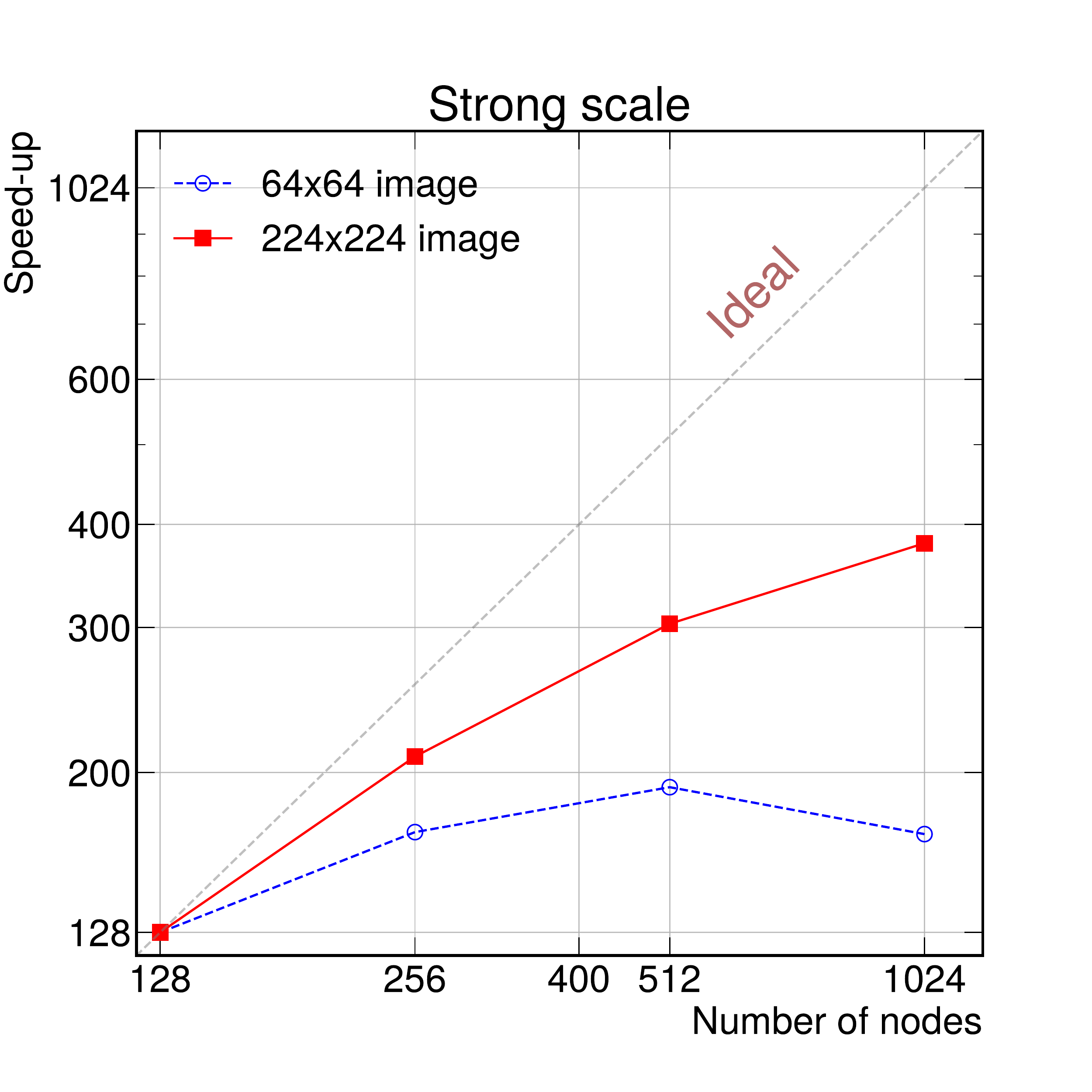}
  \label{fig:sfig1}
\end{subfigure}%
\begin{subfigure}{.5\textwidth}
  \centering
  \includegraphics[width=.9\linewidth]{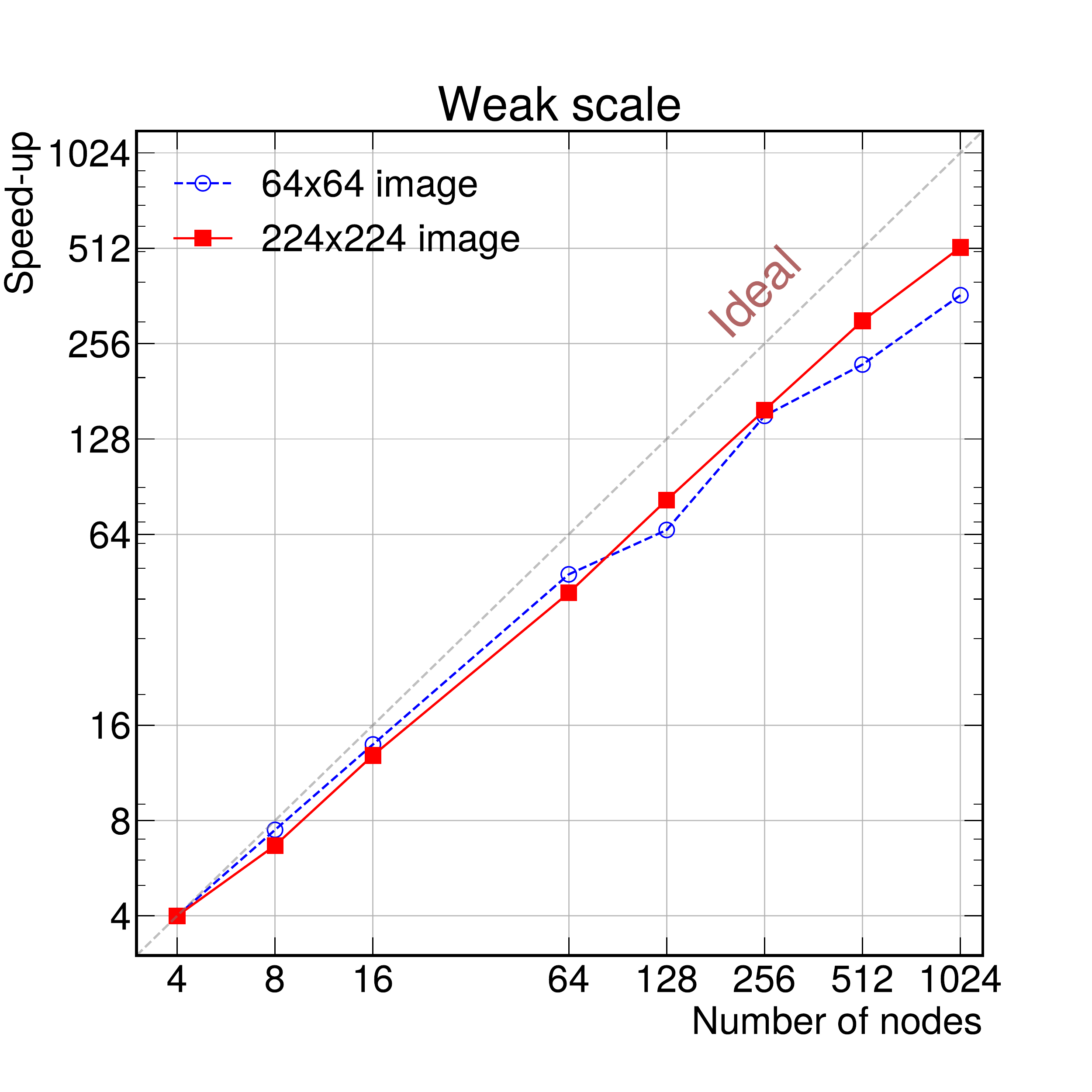}
  \label{fig:sfig1}
\end{subfigure}%
\caption{Training time per epoch according to the number of nodes (upper panel) and the speedup plots (lower panel). Scalability of $224\times224$ and $64\times64$ images are compared. The $224\times224$ images shows the scalability of up to 1024 nodes. Results from the ``weak scaling'' are included for reference.}
\label{fig:Scale}
\end{adjustbox}
\end{figure} 

In addition, the training time per epoch vs the number of nodes is shown (in the upper panel of Figure \ref{fig:Scale}). As mentioned in previous Section, all the results obtained thus far are from ``strong scaling''. However, from now on, we also introduce the results of ``weak scaling'' as a reference. In the ``strong scaling'', the training time of the $64\times64$ images decreases as the number of nodes increases up to 512 nodes. However, the training time with 1024 nodes is longer than that of 512 nodes. In this case, the communication time between the node is expected to affect more significantly to total running time than the training time. However, the results obtained using the $224\times224$ pixel images shows scalability according to all the nodes, because the time spent in training the $224\times224$ pixel images is dominant over the communication time between the nodes. In the ``weak scaling'', the training time decreases as the number of nodes increases. 
  
Finally, we compare the ``speed-up'' factor, which is equal to the training time of N (N$>$1) nodes divided by that of a single node (lower panel in Figure \ref{fig:Scale}). Ideally, the ``speed-up'' should be linear, however, the communication time between the nodes results in a curved plot. In the ``strong scaling'', the CNN model with $224\times224$ pixel images shows scalability from 128 nodes to 1024 nodes, whereas that with $64\times64$ pixel images shows scalability up to 512 nodes. In addition, the speed-up plot obtained from the $224\times224$ pixel images is closer to the linear plot than obtained from $64\times64$ pixel images. In the ``weak scaling'', both results with $224\times224$ and $64\times64$ images, which are close to linear function, show scalability. In conclusion, we verify the scalability of the CNN model using $224\times224$ pixel images, up to 1024 nodes in the Nurion HPC system.

\section{Summary}
We build a CNN model that accepts low-level information as images, considering the geometry of the CMS detector, and implement this model to discriminate the RPV SUSY signal events from the SM (QCD multi-jet) background events. The goal of this study is to compare the classification performance of the CNN method with that of the conventional cut-based method, which has been widely used in HEP to date. To decrease the training time, we perform the study using the Nurion HPC system at KISTI, which is equipped with thousands of parallel “Xeon Phi” CPUs. Therefore, the second goal of this study is to ensure that our CNN model runs efficiently in the Nurion HPC system and show scalability.
  
Both the signal and background have almost 10 jets in the final states. We interpret these final 10 jets as detector images considering the full sub-detector system, which indicates the HCAL, ECAL and tracker. We consider these three sub-detector images as a single image of three different color channels: red, green, and blue. Further, two different pixel numbers ($224\times224$ and $64\times64$) are employed in the image generation step for the same detector coverage. The $224\times224$ pixel image, which has a higher resolution, is generated, because it is expected to have more information than that in the $64\times64$ pixel image. The background samples are weighted to consider the probability of event-generation, and rescaled to a weighted sum of background events, which is equal to the number of signal events, to prevent imbalance between the signal and background data. Furthermore, we optimize the network architecture by implementing a new padding method, called ``circular padding'', considering the azimuthal symmetry of the image data obtained from the shell of a cylindrical detector.
  
Notably, the classification performance of the CNN method is found to be better than that of the cut-based method. We use approximately \num{750000} images in each of the two categories: RPV SUSY signal and QCD multi-jet background (number of data points: \num{450000} for training, \num{150000} for validation and \num{150000} for test). The signal efficiency using the CNN method is 1.845 (1.803) times higher than that obtained using the cut-based method for the $224\times224$ ($64\times64$) pixel images. We obtain an average AUC of 0.9903 (0.9914) for the $224\times224$ ($64\times64$) pixel images. In addition, we compare the expected significance of the cut-based method with that of the CNN method at the same background rejection point (FPR=0.003503), and observe that expected significance of the CNN method is 1.2 times higher than that of the cut-based method. Furthermore, we investigate the number of expected events in the discovery level (5$\sigma$), from the CNN results. There are 258 (259) signals and 2476 (2473) backgrounds, for the $224\times224$ and $64\times64$ pixel images. Evidently, the classification performance obtained from using the $224\times224$ pixel images is almost same as that obtained from using the $64\times64$ pixel images.
  
We demonstrate the scalability of our CNN model up to 1024 nodes in the Nurion HPC system, using the $224\times224$ pixel images based on the training time and ``speed-up'' factor which is calculated as the training time of the N nodes divided by that of a single node. As the number of nodes increased, the training time decreased, and the ``speed-up'' factor increased. We also compare the training time of 1024 KNL nodes with a single GPU machine (TITAN RTX). The training time per epoch using 1024 nodes is found to be 25 times faster than that observed using the GPU. 

\ack
The authors would like to thank Wahid Bhimji and Steven Farrell for helpful discussions in the initial stage of this work and for helping in the careful review of this manuscript. This research was supported by the National Research Foundation of Korea (NRF) grant funded by the Korean government (MSIT)(Grants No.2018R1A6A1A06024970, \\ No. 2020R1A2C1012322 and Contract NRF-2008-00460). This research used resources of the National Supercomputing Center and the computing resources of the Global Science experimental Data hub Center (GSDC) in Korea Institute of Science and Technology Information (KISTI), along with supercomputing resources and technical support.
\printbibliography

\end{document}